\documentclass[12pt]{revtex4}
\usepackage[dvips]{graphicx,color}
\usepackage{pslatex}
\def\bea{\begin{eqnarray}}
\def\eea{\end{eqnarray}}
\def\be{\begin{equation}}
\def\ee{\end{equation}}
\def\nn{\nonumber}

\def\m{\mu}
\def\n{\nu}

\def\om{\omega}

\def\t{\tau}

\def\e{\epsilon}

\def\be{\begin{equation}}
\def\ee{\end{equation}}
\def\ovl{\overleftarrow}
\def\ovr{\overrightarrow}

\begin{document}

\title{Quantum Field Theory in Accelerated Frames}
\author{Arundhati Dasgupta,\footnote{E-mail:arundhati.dasgupta@uleth.ca}}
\affiliation{\\ 4401 University Drive, \\ University of Lethbridge,\\
Lethbridge T1K 3M4.\\}

%\maketitle

\begin{abstract}
In this paper we re-investigate the Bogoliubov transformations which relate the Minkowski inertial vacuum to the vacuum of an accelerated observer. We implement the transformation using a non-unitary operator used in formulations of irreversible systems  by Prigogine. We derive a Lyapunov function which signifies an irreversible time flow. We extend the formalism to the black hole space-time which has similar near-horizon geometry of an accelerated observer, and in addition show that thermalization is due to presence of black hole and white hole regions.  Finally we discuss an attempt to generalize quantum field theory for accelerated frames using this new connection to Prigogine transformations.
\end{abstract}

\maketitle

\section{Introduction}

In 2015 the world celebrates 100 years of General Relativity (GR). Einstein discovered GR while trying to generalize the theory of special relativity to non-inertial and accelerating frames. In this new theory physics was invariant under `general coordinate transformations' including transformations to frames of accelerated observers. Simultaneously as GR quantum mechanics (QM) was being developed as a theory of the atomic and molecular regime. Initially the theory of QM was formulated using `Galilean' invariant Schr\"{o}dinger equation. The QM for inertial or constant velocity frames, which respected the laws of Einstein's special theory of relativity was formulated much later. 
Quantum field theory (QFT) was developed initially by Dirac \cite{qft} and then in the 1950's and 1960's by various physicists who generalized quantum mechanics for relativistically invariant systems \cite{qft1}. The obvious question which follows is what is QFT in accelerated frames, and how does one reconcile this with gravity as GR? Can this lead to the formulation of a quantum GR? This has been a notoriously difficult task \cite{qgr}, however, our perspective is that the clues for a quantum GR theory are hidden in QFT for accelerated observers. In this paper we investigate QFT for accelerated observers using a formulation in complex physics due to Prigogine \cite{prig}, and try to obtain some insights from the formulation for a quantum theory of  GR.
 
We begin by re-investigating an accelerated observer in Minkowski space-time, where the physics is well understood. An accelerated observer's frame of reference is described using the Rindler coordinates \cite{wald,marolf}. In this frame, the Minkowski causal diamond appears as causally disconnected into two Rindler universes with time flowing in the forward direction in one universe and reverse direction in the other. An accelerating observer restricted to one Rindler universe sees the Minkowski vacuum as a state with particles distributed in a thermal spectrum with a temperature proportional to his/her proper acceleration. This can be observed by obtaining the creation and annihilation operators of the Rindler QFT's as linear transforms of the Minkowski operators \cite{bd1,unrhwald}.  The transformation known as the Bogoliubov transformation can also be implemented on the vacuum states.  We formalize these transformations, and obtain them as non-unitary operators which act on the density matrices of the Minkowski space-time.  The density matrices of two Rindler Hilbert spaces (one with time flowing forward and the other with time flowing backward) are obtained using this and the individual operators of the Rindler Hilbert space are also obtained using non-Unitary transformations.  We find that these `formalized' transformations have been previously studied in physics of irreversible systems due to Prigogine \cite{prig}.

I. Prigogine \cite{prig} had reformulated complex physics, where systems are characterized by thermalization and irreversible time evolution using new `star-Hermitian' operators. These were non-unitary operators, and `star-Hermitian' is an operation which in addition to Hermitian conjugation involves time reversal. Curiously these operators were defined by implementing a linear transform from density matrices of normal `unitary'  Hilbert space to density matrices of an irreversible system. In the transformed Hilbert space the time evolution is not unitary and `entropy' production occurs due to presence of an anti-Hermitian term in the star Hermitian evolution operator. The presence of entropy and irreversibility in the transformed Hilbert space is identified by building a Lyapunov function which decreases in any physical process. We show that each of the the Rindler vacuum density matrix  is a very similar `non-unitary' operator transform of the Minkowski vacuum density matrix: the operators are non-unitary. The time evolution equation in the `Rindler' Hilbert-space is non-unitary and we construct a Lyapunov function for the Rindler observer and show that this has irreversible behavior, signifying the existence of `entropy'. However, what is interesting is that unlike the systems discussed in \cite{prig}, the transformation connects Hilbert spaces of two different coordinate systems, with two different times. Thus, if we flow the Lyapunov function in Minkowski time, there is no entropy creation. If instead we flow the Lyapunov function using the Rindler Hamiltonian, there is monotonic flow.   

We also clarify some of the criticisms of Prigogine's formalism by providing a very concrete and transparent system for its implementation. In particular, for the Rindler system, we show that the initial unitary Hilbert space is mapped to two different Hilbert spaces by two versions of the Prigogine
transformation. One Hilbert space has the time as reverse of the time of the other Hilbert space. Irreversibility arises in one Hilbert space due to the ignorance of one of the Hilbert spaces, and choosing one of the Prigogine transformation over the other, representing unidirectional time flow.

We then generalize the above formalism to the case of the black hole space-time. The near horizon region of the black hole has the same metric as that of the Rindler observer. The bifurcate horizon disconnects two asymptotics (I and II as in Figure 2), which play the same role as two Rindler universes.  The Kruskal vacuum  of the \cite{bd1} of the space-time plays the role of the Minkowski vacuum and a transformation of this to the individual I and II Rindler vacuums is derived as a Prigogine transformation. 
The black hole space-time is different from the Minkowski space-time, as it has black hole and white hole regions, where matter can propagate in one direction only, into the black hole and out of the black hole. This gives the entropy of the black hole a real existence `independent of observer' as opposed to Minkowski-Rindler example, but induced due to boundary conditions. 

Having formulated two different examples of QFT as observed by accelerating observers, 
we finally formulate a general theory of transforms of quantum density matrices of QFT from inertial to non-inertial frames, and discuss the implications of this.  

For other discussions/derivations on Rindler space-time thermodynamics see \cite{rindl,rindl2} and for discussions on accelerated frames see \cite{accel,accel2} and references therein.
In the next discussion we describe the formulation due to Prigogine for physics of complex systems and the star-Hermitian operators. We introduce the generic quantum field theory vacuum and discuss the transformation to the Rindler vacuum in the second section. In the third section we formalize these transformations as Prigogine transformations. In this section we also
construct the Lyapunov function and the time evolution which shows irreversible behavior in the Rindler frame.
The fourth section deals with accelerating observer near a black hole horizon. The fifth section generalizes QFT for accelerating frames using the connection to the Prigogine transformation. The sixth section is a conclusion. 
\subsection{The Complex Physics}
Nature is inherently complex, and systems evolve irreversibly. In physics we idealize and isolate systems and formulate reversible, unitary laws describing the dynamics. In reality synergistic behavior of systems comprised of many fundamental entities is prevalent in natural phenomena. A typical system is the `ideal gas' consisting of molecules whose individual dynamics is reversible and unitary but collectively they are `thermalized' and possess entropy. It is generally agreed that `entropy'
arises due to loss of information of the system, and irreversibility results due to `random interactions' amongst the constituents which introduces a degree of unpredictability in the system. 
The origin of `entropy' and unidirectional time evolution was initiated from Boltzmann's H-theorem \cite{boltz}. Since then variations of the H-theorem have tried to re-interpret the origin of entropy. One such attempt is due to Prigogine whose ideas tried to bring to the physics of complex systems the concept of microscopic entropy. In a very fundamental formulation of irreversible systems, he introduced operators which transformed density matrices of a typical Hilbert space into density matrices of a Hilbert space whose time evolution had irreversible behavior. In the transformed Hilbert space the time evolution equation was non-unitary \cite{prig}. 
There is a very subtle difference between the usual `thermalization process' where the irreversibility emerges in the macroscopic averaging process, and the irreversibility of the Prigogine formulation, here irreversibility is introduced using a non-unitary transform from a unitary Hilbert space in the quantum regime, and thus microscopic in origin.

In the following we give a very brief and cryptic description of the formulation. For further details refer to \cite{prig}. Let us take a system which is described using a density matrix $\rho$ and its dynamics is described using a Hamiltonian. This density matrix $\rho$ evolves in time using the Liouville operator L, obtained as the commutator of the Hamiltonian H with the density matrix. 

\be
\iota \  \hbar \frac{\partial \rho}{\partial t} = [H,\rho]= L\rho
\label{eq:liou}
\ee
This is typical in a quantum mechanical system, and if the Hamiltonian is Hermitian, the evolution is unitary. However, a non-unitary irreversible flow can be obtained by defining a `transform' from the above density matrix.
This is implemented using a non-unitary operator
$\Lambda$, which is a function of the Liouville operator L.
\be
\tilde\rho= \Lambda^{-1}(L)\rho 
\label{eqn:transformation}
\ee
The $\Lambda$ operator is such that it obeys
\be
\Lambda^{-1}(L)=\Lambda^{\dag}(-L)
\label{eqn:star}
\ee

This is motivated from the requirement that expectation values of observables remain same in both the representations of the
Hilbert space.
\be
{\rm Tr}(A^{\dag} \rho )={\rm Tr}(\tilde A^{\dag} \tilde\rho)
\ee
where the observable transforms as $\tilde A= \Lambda^{-1} (-L) A$. 
Notice, the interesting $L\rightarrow -L$ in the argument of the $\Lambda$ operator. This operation can be interpreted as the `time-reversal' operation as obviously in equation (\ref{eq:liou}) the $t\rightarrow -t$ implements $L\rightarrow -L$.
This time reversal accompanied with Hermitian conjugation defines the `Star Hermitian'  operation: 
\be
\Lambda^*=\Lambda^{\dag}(-L)
\ee
Thus we can see that we can write equation (\ref{eqn:star}) as
\be
\Lambda^{-1}= \Lambda^*.
\ee

The transformed evolution equation which evolves the transformed $\tilde \rho$ matrix is thus:
 \be
\iota \  \hbar  \frac{\partial \tilde\rho}{\partial t}= \Phi \ \tilde \rho
\ee
where $\Phi= \Lambda ^{-1} L \Lambda$.
The resultant evolution operator can be written as a star Hermitian operator as
\be
\Phi^*= \Phi^{\dag}(-L)=-\Phi(L)
\ee
Or 
\be
(\imath \  \Phi)^*= \imath \ \Phi
\ee
This `star Hermitian operator'  can be written as the sum of a Hermitian and a anti-Hermitian operator.
\be
\imath \  \Phi=\imath\  \Phi_{e}+ \imath \ \Phi_{o} \ \
\ee
The entropy production comes from the `even' part of the operator $\imath \ \Phi_e$, it can be shown that this term is similar to the collision term in Boltzmann's H-theorem. 
For further information in this field there are papers describing the Friedrich's model in condensed matter systems \cite{michel}.
\subsection{Lyapunov functions}

Lyapunov functions are well defined in mathematics \cite{lyapunov}, for studying stability of systems. They can be defined as positive definite functions whose derivatives have monotonic behaviour with asymptotic stability. In physics these are used for systems to describe entropy which `always' increases with time. Further the system acquires equilibrium in the maximum entropy state. In the Prigogine formalism,
Lyapunov functions can be defined in the transformed space e.g.
\be
\Omega= {\rm Tr}(\tilde \rho^{\dag}\tilde\rho)
\ee
using the density matrix obtained in (\ref{eqn:transformation}). 
The time derivative of this
\be
\frac{d\Omega}{dt}= \frac{d}{dt}\left[{\rm Tr}\ \rho^{\dag}(0)e^{iLt} \ T^{\dag}T \ e^{-iLt}\ \rho(0)\right]
\ee
where $T=\Lambda^{-1}$.
is non-zero, as the Liouville operator does not usually commute with $\Lambda^{\dag}\Lambda= M$ operator.
Defining $T^{\dag}T$ as the super operator $M$, one finds
\be
\frac{d\Omega}{dt}=  -{\rm Tr}\ \rho^{\dag}e^{i Lt} i \ (M\ L-L\ M)e^{-iLt}\rho(0) \leq 0\nn \\
\ee
 In \cite{prig} the  operator ML-LM=D is defined as the entropy operator and measures the `irreversibility' of the system. The significance is in the non-commuting terms (ML-LM) which gives rise to irreversible flow. 

In the next section we describe the quantum state of an accelerated observer as obtained from the Minkowski state motivated from this formulation, and obtain precisely such an irreversible flow of the Rindler vacuum.  

The existence of Lyapunov functions with asymptotic stability, give rise to entropy creation in a system \cite{prig}. 
%\subsection{Irreversibility}
%We now examine the example of the Rindler space. As the accelerated observer splits the Minkowski universe into two Rindler universes, there are two different Prigogine transformations which connect the Minkowski operators to two different vacuums. We discuss this in details next.
%\subsection{The Prigogine Operator Transforms:}
%In this section we analyze the nature of the operator, $\Lambda(L)$ which maps the density matrices of one Hilbert space, to an %inequivalent one, which shows non-Unitary evolution.
%As stated above, this process brings in an irreversibility, by preferably choosing one direction of time. Thus starting from a reversible %system, represented as direct product of two Hilbert spaces, each of which has uni-directional time behaviour the map chooses one. We will %discuss this in more details when we take the concrete example of the Rindler observer.
%Thus
%\be
%\tilde\rho=\Lambda(L) \rho
%\ee

%\be
%H = H_{\rm mapfwd}\otimes H_{\rm mapback}
%\ee
%The $H_{\rm mapback}$ is the time reversed version of $H_{\rm mapfwd}$. A transformation takes the operators of the original Hilbert %space to one of the forward/backward Hilbert spaces. Thus the transformation breaks the time symmetry.

% Is the operator Linear?
%\bea
%\Lambda(L) (\rho_1+\rho_2) &= &\Lambda(L)\rho_1 +\Lambda(L)\rho_2\\
%\Lambda(L)(\alpha \rho)&=&\alpha \Lambda(L)\rho
%\eea
%The operator is a Linear operator as it satisfies the above two criteria.

\section{Accelerated Observers}

In this section we explore the quantum field theory of an accelerated observer in Minkowski space. It is well known that there is particle creation in the frame of an accelerated observer \cite{unruh} and we examine this in the
formalism introduced in the previous section. We begin with a discussion of the quantum field in Minkowski space-time.
\subsection{Quantum Field Theory in Minkowski Space}
The Minkowski space is flat and the metric of the space-time in Cartesian coordinates is
\be
ds^2=-dt^2 + dx^2+ dy^2+ dz^2
\label{eqn:minkowski}
\ee
Most of pre-General Relativity physics and quantum mechanics as well as quantum field theory is formulated in Minkowski space.  We take the example of the simplest field: the scalar field to illustrate the transform from the inertial observer to the accelerated observer. Defining a scalar quantum field using a Lorentz invariant action of the form \cite{qft}
\be
-\frac12 \int d^4 x \ (\partial_{\mu}\phi\partial^{\mu}\phi -m^2 \phi^2)
\label{eqn:action}
\ee
where $m$ is the mass of the scalar field $\phi$ (which is taken as real) one obtains the dynamics.
The scalar data is given on a Cauchy surface and evolved in the Minkowski space-time using the appropriate equation of motion.
The inner product of the scalar fields on a hyper surface of the Minkowski space is defined as
\be
<\phi_1, \phi_2> = \int_{\Sigma} d\Sigma^{\mu} \left(\partial_{\mu}\phi_1^* \phi_2 - \phi_1^*\partial_{\mu}\phi_2\right)
\label{inner}
\ee
%The equation of motion of the scalar field obtained by varying the action (\ref{eqn:action});
%\be
%(\partial^{\mu}\partial_{\mu}- m^2)\phi=0
%\label{eq:eom}
%\ee

The quantization of the field in the Fock space representation proceeds by identifying the positive frequency modes defined as conjugate to the `time coordinate' of the background field. The Hamiltonian defined canonically using this background time measures the 
quanta of these positive frequency modes. The `creation' and `annihilation' operators for these modes defined using the canonical momenta and the position vectors `factorize' the Hamiltonian, and the field configuration is defined using these operators.  
%Using the flat metric this equation is 
%\be
%\left[-\frac{\partial ^2}{\partial t^2} + \frac{\partial^2}{\partial x^2} + \frac{\partial ^2}{\partial y^2} + \frac{\partial ^2}{\partial z^2}\rig%ht]\phi=0
%\ee

The positive frequency solutions are identified with respect to time as
\be
\frac{\partial \phi}{\partial t}=-i\omega\ \phi
\ee
Using the (\ref{inner}) inner product, one can normalize the wave function. The `mode' solution is obtained as
\be
\phi= \frac{1}{\sqrt{(2\pi)^3 \omega} }e^{- i \ \omega \  t + i \ \vec k.\vec x}
\ee

where the vector $\vec k$ is the momentum vector and the dispersion relation is
\be
\omega^2 -k^2 =m^2
\ee
We shall set $m=0$ in the rest of the paper.
 The quantum of these normal modes are created and annihilated by operators $a_{\omega, \ \vec k}^{\dag}, a_{\omega,  \ \vec k}$
which appear in the description of the Hamiltonian. 
\be
H= \int  d^3 \vec k \  \omega  :a^{\dag}_{\omega, \vec k}~ a_{\omega, \vec k}:
\ee  

A typical `quantum' configuration of the scalar field is then

\be
\Phi= \int \frac{d^3 \vec k}{(2\pi)^{3/2} \sqrt{\omega}} \left[ a_{\omega,  \vec k} \ e^{- i \ \omega \  t + i\ \vec k.\vec x} + a^{\dag}_{\omega,  \vec k} \ e^{i \ \omega \ t - i \ \vec k.\vec x}\right]
\ee

It is obvious from this that

\be
<\Phi, \phi_{\omega, \vec k}>= a_{\omega, \vec k}
\ee
and
\be
<\Phi,\phi^*_{\omega, \vec k}>= - a^{\dag}_{\omega, \vec k}
\ee

The vacuum state of the theory is a unique vacuum and it is defined as the state with no quanta. And the annihilation operator
annihilates the vacuum.
\be
a_{\omega,\  \vec k}\ |0_M>=0
\ee
The modes with energy $\omega$ and momentum vector $\vec k$ can be created out of the vacuum using $a^{\dag}_{\omega\  \vec k}$.

 These modes begin at the far past and propagate to the far future. However, for the massless modes, the Minkowski scalar fields are best described using the causal diamond. The scalar fields travel along light like geodesics from one asymptotic past to the other asymptotic future. There are two sets of asymptotics.

The light cone coordinates are defined as:

\be
v=  t+x \  \  \  \  \  u= t-x
\ee

and correspondingly we define $k_u, k_v$ as the light cone momenta coordinates. The light cone creation and annihilation operators are obtained by making the following transformations
\be
\sqrt{\omega}\  a_{\omega, \vec k}=\sqrt{ k_v} \ a_{k_v\ \  \vec {\bf k}}
\ee
 
where we have defined the Light cone mode as $e^{i\  k_v \ u + i \ k_u \ v + i \ \vec{\bf k}.\vec {\bf x}}$.
\be
k_v=  \om -k_x \  \  \   \  \  \  k_u=  \om + k_x
\ee
 
 $\vec {\bf x}$ are the coordinates perpendicular to the light cone directions. We define the above modes on a $v$ constant surface.
The v=constant and u=constant surfaces are defined at 45$^\circ$ angles to the t-x axes and the wavefronts  propagate along these.   The time reversed Minkowski modes will become important in the following discussion.
We will specifically work with the $v=0$ surface as in \cite{unrhwald} and on this surface, the scalar field will be
\be
 \Phi (u, {\bf x})=\int \frac{d k_v}{\sqrt{(2\pi)^3 k_v}}\  d^2 {\bf k} \ ( {\ovr a}_{k_v \bf k} e^{-i k_v u + i \bf k\cdot \bf x} + {\ovr a}^{ \dag}_{k_v \bf k}\  e^{ i k_v u - i \bf k\cdot \bf x} )
\ee
The time reversed version of this $\Phi (-u, \bf x)$ will also be used to describe the Rindler modes, and the corresponding annihilation and creation operators ${\ovl a}_{k_v \bf k}, \ {\ovl a}^{ \dag}_{k_v \bf k}$ of this will be defined for this purpose.

What is interesting is that to obtain the complete transformation from the Minkowski to the Rindler space, we require two sets of modes Minkowski modes; 
$\ovr{a}_{k_v \  \bf k}$ and $\ovl{a}_{k_v \  \bf k}$. This interpretation of the modes as creator of forward and backward light cone travelers serves the purpose of the Prigogine formalism.
The light cone Hamiltonian is
\be
\int d^2 {\bf k} \ dk_{v} \ \  k_v \  :a^{\dag}_{ k_v\   \bf k} a_{k_v\   \bf k}:
\ee

\subsection{Quantum Field Theory for an Accelerated Observer} 
The accelerated observer in flat space-time describes a Rindler metric. This metric is obtained using a coordinate transformation which takes us to the frame of the accelerated observer. The coordinate lines are actually trajectories of `boost isometries' of Minkowski space generated by $b^a = x\frac{\partial}{\partial t}^a + t \frac{\partial}{\partial x}^a$. Given the Minkowski space-time (\ref{eqn:minkowski}), the coordinates along the trajectories of the boost Killing vector are:

\be
\tau= \frac{1}{2 a}  \tanh^{-1}\left(\frac tx\right) \  \  \  \  \  \   \   \   \   \   \   \   \  \rho=  \frac1{4a} (x^2- t^2)
\label{eqn:boost}
\ee 
 
The Rindler metric is
\be
ds^2=-\frac{\rho}{a} \ d\t^2 + \frac{a}{\rho} \ d\rho^2   + dy^2 + dz^2
\label{eqn:metric}
\ee

The Rindler observer sees only wedges of the Minkowski space. These are described as $x>|t|, x<-|t|$ regions of the Minkowski space-time.
 
\begin{figure}
\includegraphics[scale=0.7]{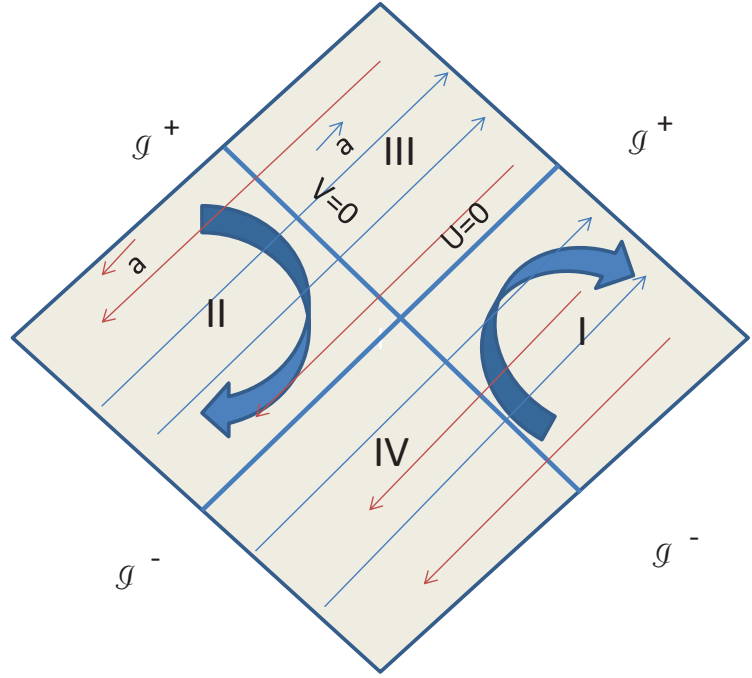}
\caption{Rindler Wedge}
\end{figure}

The $\rho$ constant Rindler observers describe hyperbolic paths $x^2-t^2=$ constant paths in Minkowski space and are the world-lines of accelerating observers.
Given the above metric, we describe the quantum field as observed by the Rindler observer.

As we described in Equations (\ref{eqn:boost},\ref{eqn:metric}), an accelerating observer does not observe the entire space-time. His/her view is
obscured by the presence of Rindler horizons, the $t=\pm x$ axes of the Minkowski space-time. One can still define quantum field theory in the two wedges which comprise the Rindler space-time, separately. 
The scalar equation in Rindler space-time appears as:
\be
-\frac{a}{\rho} \partial_{\tau}^2 \phi + \partial_{\rho}\left(\frac{\rho}{a}\partial_{\rho}\phi\right) + \partial^2_y \phi+ \partial^2_z\phi=0
\ee
The positive frequency modes are defined wrt Rindler time $\tau$. These are therefore
\be
\frac{\partial \tilde\phi_{\om}^R}{\partial \tau}= -\imath \ \om \ \tilde \phi^R_{\om}
\ee

The ansatz for the wave function is $\tilde\phi^{R}_{\om}= e^{-\imath (\omega \tau+ k_z z + k_y y)} \chi(\rho)$, where $\chi(\rho)$ is the solution of the equation:
\be
\rho^2\  d_{\rho}^2 \chi + \rho\  d_{\rho}\chi + \left[ a^2 \omega^2 - a \ \rho (k_y^2+ k_z^2)\right] \chi=0
 \ee
 This equation can be solved in terms of Bessel functions and the solutions are regular at $\rho=0$.  However the solutions are valid in either the Rindler wedge I or II  (see Figure 1).    
The solution is obtained as
\be
\chi_I(\rho)= A_I I_{2 i \om \  a} (2 k \sqrt {\rho}) +  B_I K_{2 i \om \ a} (2 k \sqrt{\rho})
\label{eqn:sol}
\ee

Where $I_{\nu}$ and   $K_{\nu}$ are the Bessel functions of the second kind. $k=\sqrt{a} |{\bf k}|=\sqrt{k_z^2 + k_y^2}$. 
 These solutions represent the solutions in Rindler wedge I and for solutions in Rindler wedge II, we use similar notations except $I\rightarrow II$.                                               
A  scalar field expansion in Rindler time gives
\be
\Phi= \int d \omega \ [ a_{ \omega \  {\bf k} \  I/II }\  \phi^R_{\omega \  I/II} +  a_{\omega, \  {\bf  k} \  I/II}^{ \ \dag} \phi^{R *}_{\omega \ I/II}]
\ee
where $\phi^R_{\om I}$ are the modes in region I and $\phi^R_{\om II}$ are the modes in region II and $a_{\om {\bf  k} I/II}$ are the annihilation operators of region I and II. There are two different vacuums 
\be
a_{ \ \om \ {\bf k} \ I} \  |0_{ R I}>=0 \  \  \  \   \  \  \  \  \  \  \  \  a_{  \om \ {\bf k} || } \ |0_{ R II}>=0
\ee
Further more the time coordinate in Rindler wedge II runs backward on the $\rho=$ constant surfaces. This can be seen using the equations for 
\be
e^{\tau/a}=\left|\frac{v}{u}\right|
\ee

and observing the derivative of $\tau$ wrt to $u$.
\be 
e^{\tau/a} \ \frac{1}{a}\frac{d\tau}{du}= \frac{v}{u^2}
\ee

In region I $v$ is positive, therefore the derivative is positive and $\tau$ increases with $u$. In region II $v$ is negative and thus $\tau$ decreases with $u$. 
Thus time in one wedge is negative of the time in the other wedge. Motivated from \cite{unrhwald}, we use the v=0 surface as an almost Cauchy surface to analyse the transformations from Rindler to Minkowski modes.
At the $v=0$ surface from (\ref{eqn:boost}), the coordinate $\rho=0$, and thus one uses the approximation to the Bessel functions around $\rho=0$. On the $v=0$ surface as $-\infty <u <\infty$, the scalar fields from both the Rindler wedges I and II contribute to the scalar data on the surface. For $u>0$ the Rindler modes for wedge II are analytically continued into the $v=0$ surface, similarly for $u<0$. Thus the field
\bea
\Phi &= & \int d\omega \  \ d^2 {\bf k} \  [ \  e^{-i \omega \tau'} e^{i \bf k \cdot x}  \chi (\rho) \ a_{\om {\bf k} II} + e^{i \om \tau' }e^{- i \bf k\cdot x}\chi^*(\rho) \ a^{\dag}_{\om {\bf k} II} \  ]  \  \  \ \     u>0 \\
 &=& \int d\om \ \ d^2 {\bf k} \  [ e^{-i \om \tau} e^{i \bf k \cdot x} \chi (\rho) \ a_{\om {\bf k} I} + e^{i \om \tau} e^{- i \bf k\cdot x}\chi^*(\rho) \  a^{\dag}_{\om {\bf k} I}
\ ] \     \  \   \  \  \  \   u<0
\eea
The time in Rindler wedge I can be written as $\tau'=-\tau$. The solution in the scalar sector is approximated near $\rho=0$ using the Bessel function (\ref{eqn:sol}) as
\be
\chi (\rho) \approx A_{I(II)}\  e^{i \om \  a \ln \rho} + B_{I(II)}\  e^{-i \om\  a \ln \rho}
\label{eqn:near}
\ee
 Plugging this in gives:
\bea
\Phi &= &\int \ d\om \ d^2 {\bf k} \left[\{A_{II} e^{\ i\om \ (\tau+ a \ln \rho)} + B_{II} e^{\ i\om \ (\tau- a \ln \rho)}\}e^{i \bf k\cdot\bf x} \ a_{\om {\bf k} II} \right. \nn \\ & +& \left. \{A^*_{II} e^{-\ i\om \ (\tau+ a \ln\rho)} + B^*_{II} e^{-i\ \om \ (\tau- a \ln\rho)}\}e^{- i \bf k\cdot x} \ a_{\om {\bf k}II}^{\dag} \ \right]
\eea
This is for $u>0$. For the other half of the v-axis, one uses
\bea
\Phi &=& \int\  d\om \  d^2 {\bf k}\  \left[\ \{ A_I e^{-i \om \ (\tau- a\ \ln\rho)} + B_I e^{i \om \ (\tau+ a\ \ln \rho)}\}e^{i \bf k \cdot \bf x}  \ a_{\om {\bf k} I}  \right.   \nn \\&+&\left.
\{ A^*_{I} e^{i \om \ (\tau- a \ln \rho)} +  B^*_{I} e^{-i\om \ (\tau+ a \ \ln \rho)}\} e^{-i \bf k.x}  \ a^{\dag}_{\om {\bf k} I}\ \right]
\eea
As it is obvious that $\tau- a \ \ln\rho=a\ln(4a \  u^2)$ and $\tau+ a\ \ln\rho= a \ln (4 a \ v^2)$, one obtains ignoring the $v$ = constant terms
\be
\Phi= \int d\om \ d^2 {\bf k} \left[ \tilde B_{II} \left(u \right)^{2i \om \  a } e^{i {\bf k}\cdot x} \ a_{\om {\bf k} II} + \tilde B_{II}^*\left(u\right)^{-2i\om \ a }e^{-i{\bf k}\cdot x} \ a_{\om {\bf k} II}^{\dag}\right]
\ee
$u>0$ and
\be
\Phi= \int d \om \ d^2 {\bf k} \left [A_I \left(u\right)^{-2i\om \ a }  e^{ i \bf k \cdot x} a_{\om {\bf k} I} + A_{I}^* \left( u \right)^{2 i \om \ a }  e^{- i \bf k \cdot x} \ a_{\om {\bf k} I}^{\dag}\right]
\ee
for $u<0$. 
The above can be Fourier transformed to the Minkowski modes $e^{-i k_v u} e^{i {\bf k}\cdot x}$. 

This therefore gets on the v=0 surface:
\bea
\Phi &= &\int d\om  \ d k_v \  d^2 {\bf k} \left[ A_{\om k_v}\left (a_{\om {\bf k} I} -  e^{-2\pi \om \ a } a^{\dag}_{\om -{\bf k} II}\right) e^{-i k_v u+ i {\bf k} x} \right.\nn \\ &+& \left. B_{\om k_v} \left( a_{\om -{\bf k} II}  - e^{-2\pi\om \ a } a^{\dag}_{\om {\bf k} I}\right) e^{-i k_v u -i {\bf k} x}\right] + c.c
\eea

where
\bea
A_{\om k_v} & \ = \ & 2^{-i \ \om \ a } \frac{\Gamma(1+ 2i\om \ a +\e)}{(-i k_v)^{1+2i\om \ a +\e}} A_I\\
B_{\om k_v}& \ = \ & 2^{i \ \om\  a }\frac{\Gamma(1-2i\om\ a -\e)}{(ik_v)^{1-2i\om \ a -\e}} B_I
\eea

(we assumed that $A_I=B_{II}^*$ etc)
The $\e$ makes the Fourier integrals well defined.
These transformations have been explicitly worked out in several other papers \cite{thooft}, this is a re-working of those to fix notations etc.  It is obvious from the Fourier transforms that the above represent `positive' frequency modes and the operators annihilation of these modes. If we define these as two new Minkowski operators:

\bea
\overrightarrow{a}_{\om \  {\bf k} }& = & a_{\om {\bf k} \ I} - e^{-2\pi \om \ a } a^{\dag}_{\om -{\bf k} \ II}\\
\overleftarrow{a}_{\om \ {\bf k}}&=& a_{\om -{\bf k} \ II} - e^{-2\pi\om \ a} a^{\dag}_{\om  \ {\bf k}\  I}
\eea

The $\ovl{a}$ annihilates a `wave packet' of positive frequency modes traveling in the positive $u$ direction and $\ovr{a}$ annihilates a wave packet  `negative' frequency mode traveling in the reverse time direction $-u$. This is the most interesting aspect of the Rindler transformations, which will be used for the Prigogine formalism. The Rindler modes are defined with both $\ovl{a}$ and $\ovr{a}$ operators, and this therefore sets the stage for the use of the formalism.
The Minkowski vacuum is written as a direct product of two vacuums,  annihilated by the $\ovr{a}$ and the $\ovl{a}$ operators. Further, a normalization has to be achieved to maintain the canonical algebra $[\ovr{a},\ovr{a}^{\dag}]=1$, and this makes the operators scale as $1/(1-e^{-4\pi\om \ a})^{1/2}$. The inverse transformation is easy to obtain: 
\bea
a_{\om \ {\bf k} I} & =& \frac{1}{\sqrt{1-e^{-4\pi\om \ a }}}\left[\ovr{a}_{\om \ {\bf k}} + e^{-2\pi\om \ a }\ \ovl{a}_{\om \ {\bf k}}^{\dag}\right] \label{eqn:trans} \\ 
a_{  \om \ {-\bf k} II }&=& \frac{1}{\sqrt{1-e^{-4\pi\om \ a }}}\left[\ovl{a}_{\om \ {\bf k}}+ e^{-2\pi\om \ a } \ \ovr{a}^{\dag}_{\om \ {\bf k}}\right]
\label{eqn:trans1}
\eea
\section{The Bogoliubov Transformation and the Prigogine Formalism}
We implement the above Bogoliubov transformation implemented on the operators on the vacuum state. As the Rindler vacuum is annihilated by the Rindler annihilation operator;   
\bea
a_{\om \ {\bf k} I }\ |0_{RI}>&=&0 \nn \\
\ovr{a}_{\om \ {\bf k}} + e^{-2 \pi\om \ a}\ \ovl{a}_{\om \ {\bf k}}^{\dag} \ |0_{RI}>&= & 0 \  \  \  \ {\rm from \ (\ref{eqn:trans})}.
\label{eqn:trans1}
\eea
The last can be used as an equation in Minkowski space to solve for the Rindler  vacuum state. It is indeed easy to see that 
\be
|0_{RI}>= \prod_{\om {\bf k}} e^{-z \ovr{a}_{\om \ {\bf k}}^{\dag}\ovl {a}_{\om \ {\bf k}}^{\dag}}|0_M>
\label{eqn:coh}
\ee
where we have used $z= e^{-2 \pi \om \ a}$. This equation is derived from Equation (\ref{eqn:trans1}), as the $|0_{R I}>$ can be interpreted as a eigenstate of the Minkowski annihilation operator. The solution in Equation (\ref{eqn:coh}) is very similar to what one might obtain for a harmonic oscillator coherent state, an eigenstate of the annihilation operator. If one would write the reverse transformation; Minkwoski vacuum in terms of the Rindler wedge vacuums, then one would write the Minkowski state as a transform implemented on the direct product of the two Rindler wedge Hilbert space states. Note that the Minkowski vaccuum as it is written in Equation (\ref{eqn:coh}) is a direct product state for the ingoing and outgoing modes of the Minkowski metric, as in Equation  (\ref{eqn:minvacc}).

This thus defines the `transformation' which we represent using the operator $U$ and its Hermitian conjugate will be the $U^{\dag}=\prod_{\om {\bf k}} e^{-z\ovl{a}_{\om {\bf k}} \ovr{a}_{\om {\bf k}}}$. This obviously is not-Unitary as $U^{\dag}\neq U^{-1}$.  

We then define a density matrix and identify the operator which implements the Bogoliubov transformation on the same.

\be
\rho_R=|0_{RI}><0_{RI}|=U^{\dag}|0_{M}><0_{M}|U
\ee 
and the transformation is thus re-writable as an operator such that:
\be
\Lambda^{-1} \ \rho= U^{\dag}\ \rho \ U=\rho_R
\ee

Define, 
\be
H= \int  \ d\om \  d^2 {\bf k}\  \om \ \ovr{a}^{\dag}_{\om \ \bf{k}}\ \ovr{a}_{\om\ \bf{k}}
\ee
Note that this represents the usual Minkowski Hamiltonian, made up of the forward modes only. 
The Liouville operator is obviously $L=[H,]$. The $\Lambda$ operator can be written in terms of the Liouville operator using
\be
U=\prod_{\om} e^{- \ L \ x}
\ee
where $$x=\frac{ z}{\omega}\ovr{a}^{\dag}_{\om \  \bf k} \ovl{a}^{\dag}_{\om \ \bf k},$$ as $$[H,\frac{z}{\omega}\ovr{a}_{\om\  {\bf k}}^{\dag}]\ovl{a}_{\om\  \bf k}^{\dag}= z\ovr{a}_{\om \ \bf k}^{\dag}\ovl{a}_{\om\  \bf k}^{\dag}. $$
One can show that 
\be
\Lambda^{-1}(L)= \Lambda^{\dag}(-L)
\label{eqn:inv}
\ee

as
\be
(\Lambda^{- 1})^{\dag} \rho= U^{\dag}~\rho ~U=\Lambda^{- 1}~\rho
\ee

and
 
\be
\Lambda^{-1}\Lambda \ \rho= \prod_{\om}e^{-(Lx)^{\dag}}\prod_{\om'}e^{(Lx)^{\dag}}~\rho ~\prod_{\om}e^{Lx}\prod_{\om'}e^{-Lx}
=\rho 
\ee
The infinite products donot complicate the algebra as the operators commute for different $\omega$.
The operator inverse is simply obtained by taking $L\rightarrow-L$, and the fact that $\Lambda^{\dag}=\Lambda$ proves (\ref{eqn:inv}) .
Operators from the Minkowski space can be transformed to the operators of Rindler Wedges I and II using the same operators e.g.
\be
a_{\om {\bf k} I}= U~\ovr a_{\om \bf{k}} ~U^{-1}
\label{eqn:opert1}
\ee
and 
\be
a_{\om {-\bf k} II}= U~\ovl{a}_{\om \bf{k}} ~U^{-1}
\label{eqn:opert2}
\ee
The conjugate operators are similarly obtained. 
The entropy operator $D$ is obtained using
\be
D=ML-LM
\ee
where $M=\Lambda^{\dag}~\Lambda$.

To check the time evolution equation in the transformed reference frame, we observe, that the usual Liouville operator has to be formulated differently when the time evolution generator, the Hamiltonian, is not Hermitian.
Let us take a density matrix
\be
\rho= |0><0|
\ee
The density matrix evolves as
\be
\rho(t)= e^{-\frac{i}{\hbar}\ H \  t} |0><0| e^{ \frac{i}{\hbar} \  H^{\dag}\  t}
\ee
If we take the time derivative of this equation we get:
\be
\frac{\partial \rho}{\partial t}= - \frac{i}{\hbar}\   H \  \rho + \frac{i}{\hbar}\   \rho\    H^{\dag}
\ee

On the above modified Liouville we use the transformation $\rho_{R}= \Lambda^{-1}\rho= U \rho U^{\dag}\rightarrow U^{-1}\rho_{R}( U^{ \dag})^{-1}=\rho$. This gives 
\be
i \ \hbar\frac{d \rho_R}{dt}= \left(UHU^{-1}\right) \rho_R - \rho_R \left(U H U^{-1}\right)^{\dag}.
\ee
Using the explicit expression for the $U$ operators and the transformation laws (\ref{eqn:opert1},\ref{eqn:opert2}), one gets:
\be
i \ \hbar\ \frac{d \rho_R}{dt}=  [H_{R I}, \rho_R] -  \int  d^2 {\bf k }\  d\omega\   \omega \  z\  \left(\ a^{\dag}_{ \om {\bf k} I} a^{\dag}_{\om {-\bf k} II}\  \rho_R -  \rho_R \ a_{ \om {\bf k} I} a_{\om -{\bf k} II}\right) 
\label{eqn:timev}
\ee

The equation has a Hermitian term and an anti-Hermitian term as expected from the formalism.

However to verify an irreversible dynamics, we build a Lyapunov function. This is of the form
\be
\Omega= {\rm Tr}(\rho_0^{\dag}T^{\dag}T\rho_0)
\ee
In this particular example $\rho_0^{\dag}=\rho_0=|0_M><0_M|$. Thus 
\be
\Omega= {\rm Tr}\ (U \rho_0 (t) U^{\dag} U\rho_0 (t) U^{\dag})
\ee

As
\be
\rho_0(t)= e^{-\frac{i}{\hbar}~ H \ t} |0_M><0_M| e^{\frac{i}{\hbar}~ H \ t}
\ee

Thus the time derivative of the Lyapunov function gives (setting $\hbar=1$)
\be
\frac{d\Omega}{dt}= i \ {\rm Tr} \ (U\rho_0(t) [H, U^{\dag} U]\rho_0(t) U^{\dag})
\ee
The commutator
\be
[H, U^{\dag} U]=[H,U^{\dag}]U + U^{\dag}[H,U]
\ee
The commutator after some algebra gives:
\be
[H, U^{\dag}] =\int d\om \ d^2 {\bf k} \om \ [\ovr{a}^{\dag}_{\om {\bf k}}\ovr{a}_{\om {\bf k}}, \prod_{\om' {\bf k}} e^{-z \ovr{a}_{\om' {\bf k}}\ovl{a}_{\om' {\bf k}}}]
\ee
As each mode is independent, 
\be
=  \prod_{\om' {\bf k}}e^{-z \ovr{a}_{\om {\bf k}} \ovl{a}_{\om {\bf k}}} \om z {\ovl{a}_{\om {\bf k}}}\ovr{a}_{\om {\bf k}}. 
\ee

And similarly for the other commutator. Thus the trace gives
\bea
\frac{d\Omega}{dt}& =& 
i <0_M| e^{i\ H \ t} \prod_{\om {\bf k}} e^{-z\  \ovr{a_{\om}}\ \ovl{a_{\om}}} \ovr{a_{\om}}\ \ovl{a_{\om}}\om z \ U - U^{\dag}\prod_{\om {\bf k}}\ovr{a}_{\om}^{\dag}\ovl{a}_{\om}^{\dag} \om z e^{-z \ \ovr{a}_{\om}^{\dag}\ovl{a}_{\om}^{\dag} }e^{-i\  H \  t} |0_M><0_M|U^{\dag} U|0_M> \nn \\
\label{eqn:mink}
\eea

As the expectation values of the operators in the Minkowski vacuum give 0.

Thus even though the $[H,U^{\dag}U]$ commutator is non-zero, there is no Lyapunov function evolution. This is not surprising as the state was evolved in Minkowski time, and the Minkowski observer being an inertial observer sees no `entropy production'. 
However, if the same state is evolved in Rindler time, using the Rindler observer or the accelerated observer's Hamiltonian, the story changes. 
Thus, the Minkowski vacuum is evolved using the Rindler Hamiltonian in time. i.e.
$|0_M(\t)>= e^{-i H_{RI} \ \t}|0_M(t=0)>$, and the Rindler state is obtained using
$|0_{RI} (\t)>= U |0_M(\t)>$. 
The same computation as in (\ref{eqn:mink}) gives a non-zero result for the derivative of the Lyapunov function in Rindler time.
\be
\frac{d\Omega_R}{d \t}=  i <0_M|e^{i \ H_{RI} \t}[H_{RI}, U^{\dag} U]e^{-i\  H_{RI} \t}|0_M><0_{RI}|0_{RI}>
\ee
The Rindler Hamiltonian can be written in terms of the Minkowski creation and annihilation operators:
\bea
H_{RI} &= & \int \  d\om \ d^2 {\bf k} \  \om \  a_{ \om {\bf k} I }^{\dag} \  a_{ \om {\bf k} I}  \label{eqn:hamil} \\
&=& \int d\om \ d^2{\bf k} \ \om \ \left( \ovr a^{\dag}_{\om {\bf k}} \ovr {a}_{\om {\bf k}} + z \ovl{a}_{\om {\bf k}}\ovr{a}_{\om {\bf k}} + z \ovr{a}^{\dag}_{\om {\bf k}}\ovl{a}^{\dag}_{\om {\bf k}} + z^2 \ovl{a}_{\om {\bf k}}\  \ovl{a}^{\dag}_{\om {\bf k}}\right) \nn 
\eea
 
The commutator for one mode of $U^{\dag} U$ is computed for brevity, and it gives:
\be
[H_{R I}, U^{\dag} U]= \omega \ \  z \ \ovr {a}_{\om {\bf k}} \ovl {a}_{\om {\bf k}} \ U^{\dag}_{\om} U - \om \  z \  \ U^{\dag} U_{\om} \ \ovr{a_{\om {\bf k}}}^{\dag} \ \ovl{a_{\om {\bf k}}}^{\dag}
\ee
The expectation value of this in the evolved Minkowski state is then found to be non-zero. This is not surprising as the function is defined in one of the Rindler sectors and is evolved in Rindler time. The other Rindler space-time is ignored. This signifies a thermalization, and from this one can postulate that the acceleration and irreversible physics are equivalent in QFT.  

We describe in details the time evolution of the Lyapunov function as follows: 
\bea
\frac{d \Omega_R}{d\tau} &=& i <0_M| \ e^{i  \ H_{R I}\  \tau}\ \left( \prod_{\om} \omega \  z \ \ovr {a}_{\om \ {\bf k}} \ovl {a}_{\om {\bf k}} U^{\dag}_{\om} \  U- \ U^{\dag}\prod_{\om} z \  \om  \ U_{\om} \  \ovr{a}_{\om \ {\bf k}}^{\dag} \ \ovl{a}_{\om {\bf k}}^{\dag} \right)\ e^{-i \ H_{RI} \ \tau} \  |0_M> \nn \\
&=& -2\  {\rm Im} <0_M| \ e^{i  \ H_{RI} \  \tau}\  \prod_{\om} \omega \  z \ \ovr {a}_{\om {\ \bf k}} \ovl {a}_{\om {\bf k}} \  U^{\dag}_{\om} \  U e^{-i \ H_{R I} \ \tau}  \ | 0_M>
\label{eqn:midn}
\eea

 if we expand the exponentials, then the odd powers of $\tau$ contribute to the Imaginary values. As the calculations become increasingly difficult, we show the computation to first order in $\tau$. 

This first order in $\tau$ of (\ref{eqn:midn}) is obtained by expanding the 
exponentials on both sides of the brackets:

 \be
 i \tau\ <0_M| H_ {R I}\ \prod_{\om} \omega \  z \ \ovr {a}_{\om {\ \bf k}} \ovl {a}_{\om {\bf k}} \  U^{\dag}_{\om} \  U |0_M> - i \tau\  <0_M| \prod_{\om} \omega \  z \  \ovr {a}_{\om {\ \bf k}} \ovl {a}_{\om {\bf k}} \  U^{\dag}_{\om} \  U \ H_{RI} |0_M>
 \ee
Using the expression from (\ref{eqn:hamil})for the Hamiltonians:
\bea
 i \tau\ <0_M| \int \  d \omega' d^2\  {\bf k'}  \ \om'  z'  \ovr{a}_{\om' \bf k'}\ovl{a}_{\om' \bf k'}\ \prod_{\om} \omega \  z \ \ovr {a}_{\om {\ \bf k}} \ovl {a}_{\om {\bf k}} \  U^{\dag}_{\om} \  U \ |0_M> && \nn \\ - i \tau  \ <0_M| \prod_{\om} \omega \  z \ \ovr {a}_{\om {\ \bf k}} \ovl {a}_{\om {\bf k}} \  U^{\dag}_{\om} \  U 
 \int d \omega' d^2{\bf k'} \  \omega' z' \  \ovr{a}_{\om' \bf k'}^{\dag} \  \ovl{a}_{\om' \bf k'}^{\dag} |0_M>&&
\label{eqn:mid}
 \eea
 
 The next set of calculations to order $\tau$ are quite straightforward, though slightly extended: Lets take the first term in the above equation (\ref{eqn:mid})
 \bea
& = &\int d\om' \ d^2 {\bf k '}\  \om' z' <0_M| \ovr{a}_{\om' {\bf k'}} \ovl{a}_{\om' \bf k'}  \prod_{\om} \om \ z \ovr{a}_{\om \bf k} \ovl {a}_{\om \bf k} \sum_n \frac{\left(- z \ovr{a}_{\om \ \bf k} \ovl{a}_{\om\  \bf k}\right)^n}{n!} \nn \\ && \prod_{\tilde\om}\sum _m 
\frac{\left(-z {\ovr{a}^{\dag}_{\tilde\om \  {\tilde{\bf k}}}}{ \ovl{a}^{\dag}_{\tilde\om{\tilde{\bf k}}}}\right)^m}{m!}|0_M>
 \eea
 where we have expanded the exponentials of the $U^{\dag}_{\om}$ and the $U_{\om}$ operators.
 This is then re-written as (essentially a normal ordering problem)
 \bea
=\int d \om  \ d^2 {\bf k} \prod_{\tilde \om} ( \om \ z)^2 \sum_n \sum_m \frac{\left(-z \right)^n }{n!} \frac{(-\tilde z)^m}{m !}  <0_M| (\ovr{a}_{\om \bf k} \ovl{a}_{\om \bf k})^{n+2} (\ovr{a}^{\dag}_{\tilde \om{\tilde{ \bf k}}} \ovl{a}^{\dag}_{\tilde \om {\tilde {\bf k}}})^m|0_M> \delta_{\om \om'}+ &&  \nn \\
\int d\omega'  d^2 {\bf k' } (\om'\  z') \prod_{\om}\prod_{\tilde \om} (\omega \ z)\sum_n \sum_m \frac{\left(-z\right)^n}{n!}\frac{\left(-\tilde z\right)^m}{m!} <\ovr{1}_{\om'}\ovl{1}_{\om'}| (\ovr{a}_{\om \bf k} \ovl{a}_{\om \bf k})^{n+1} ({\ovr{a}^{\dag}_{\tilde \om \tilde{\bf k}}} {\ovl{a}_{\tilde \om \tilde{\bf k}}} )^m |0_M> &&   (\omega'\neq \omega) \nn
\eea

 We then use the harmonic oscillator normalization $a^{\dag} |n>= \sqrt{n+1} |n+1>$ and $a |n>=\sqrt{n}|n-1>$ and orthogonality $<n_{\om}| m_{\om'}> =\delta_{\om \om'} \delta_{n m}$  to find that the above can be simplified to:
 \bea
= \int d\om d^2 {\bf k}  \prod_{\tilde \om} (\om\  z)^2 \sum_n \sum_m \frac{(-z)^n}{n!}\frac{(-\tilde z)^m}{m!} \ \delta_{\om \ \om'} \delta_{\om \tilde\om}\ \delta_{n+2, \ m}  (n+2)! m! && \\ + \int d\om' d^2 {\bf k} \prod_{\om'} \prod_{\tilde\om}(\om'\  z') (\om z)\sum_n \sum_m \frac{(-z)^n}{n!}\frac{(-\tilde z)^m}{m!}  \delta_{\omega' \tilde \omega }\ \delta_{\om \  \tilde \om} \ \delta_{n+1, \ m-1} (n+1)! m! && (\om \neq \om') \nn
 \eea
 Due to the delta functions one obtains
 \be
 = \prod_{\om} (\om z)^2 \sum_{n} (-z)^{2n+2} (n+2)(n+1) 
\ee
as the second term goes to zero as the delta functions and the condition $\om \neq \om'$ are in conflict with each other.
This infinite sum can be evaluated using infinite series summation as $z^2 <1$.

we find
\be
= \prod_{\om} \frac{ (\omega z)^2 2 z^2}{(1-z^2)^3}
\ee

where we have used:
\bea
\sum_{k=0}^{\infty} q^k & = & \frac{1}{1-q}\\
\sum_{k=0}^{\infty} k q^k  &=& \frac{q}{(1-q)^2}\\
\sum_{k=0}^{\infty} k^2 q^k &=& \frac{q + q^2}{(1-q)^3}\\
%\sum_{k=0}^{\infty} k^3 q^k &=& \frac{q + 4 q^2 + 2 q^3}{(1-q)^4}
\eea

Similarly the second term in (\ref{eqn:mid}) can be evaluated in the same way and the result is
\be
= (\omega z)^2 \sum (-z)^{2n} (n+1)^2= \prod_{\om} \frac{(\om z)^2 (1+z^2)}{(1-z^2)^3}
\ee

The final expression for (\ref{eqn:mid}) is:
\be
i \t \ \prod_{\om} (\omega z)^2\  \frac{z^2 -1}{(1-z^2)^3}= - i \t \prod_{\om} \frac{(\om z)^2}{(1-z^2)^2}
\ee

If we plug this in the, answer, the Lyapunov function increases with time!
\be
\frac{d \Omega}{d\t}= \prod_{\om} \frac{(\omega z)^2}{(1-z^2)^2}\tau
\ee

What might we have interpreted wrongly? The answer is in the fact that unlike the Prigogine discussion which talks about Galilean invariant quantum mechanics, here we are discussing Relativistic quantum field theory, in Lorentzian signature. To discuss the thermodynamic aspect of this system as in previous discussions of Rindler entropy, we have to use Euclidean time. This defines the entropy flow correctly:
setting $\t\rightarrow iT_E$
\bea
\frac{d \Omega_R}{d(iT_E)} &= & i <0_M| e^{i H_{RI} (i T_E) }[H_{RI}, U^{\dag} U] e^{- i H_{RI} (iT_E)}|0_M>\\
\frac{d \Omega} {dT_E} &=& - \prod_{\om} \frac{(\om z)^2}{(1-z^2)^2} T_E
\eea

As this does not fall into any of the usual infinite products, we take log of the expression and compute the integral in $\omega$, and then exponentiate the result. The result is a function of some cut off frequency $\lambda$
\be
\frac{d\Omega_R}{d T_E} = - T_E \ (\lambda^2)^{\lambda} e^{- 2\lambda - 2 \pi \lambda^2 a+ C} \leq 0 
\ee

The function obviously decreases with time, and like all field theoretic quantities was obtained using a regularization.
Thus there is a non-zero Lyapunov function in the Rindler observers frame and entropy production occurs evident in Euclidean time. As $a\rightarrow \infty$, the zero acceleration limit, the Lyapunov function drops out. 

It is known that for correlation functions, Euclideanisation of the time coordinate is required to obtain the thermal character of the field \cite{bd1}.

Though there is somewhat strange falacious argument, given that if we had
\be
 e^{- i H_{R I} \tau} |0_{RI}>= |0_{RI}(\tau)>
\ee
instead of
\be
U e^{-i H_{RI} \tau} |0_M>= |0_{RI} (\tau)>
\ee

the system would have evolved unitarily. However, we should interpret the Lyapunov function as
\be
\Omega_{R}= {\rm Tr} (\rho_0 ^{\dag} \Lambda^{-1 \dag}\Lambda^{-1} \rho_0)
\ee
with $\rho_0 = |0_M><0_M|$ and this has an asymptotic flow.

\subsection{ Which Time?}

The Rindler time and Minkowski time both can be used to describe the time evolution of quantum states. This therefore brings us to one of the most fundamental problems in quantization in accelerated frames, or in curved space-time: if there is no unique time, what should be the physical observable to describe the flow of the system? At this level of the discussion, we donot obtain a time invariant physics, instead we formulate the physics of the accelerated observer in the time of the observers frame and postulate:

{\it  An accelerating observer detects entropic behavior using a Lyapunov function built in his frame of reference. } 
Defining 
\be
\Omega= {\rm Tr} (\rho^{\dag}(\Lambda^{-1})^{\dag}\Lambda^{-1} \rho)
\ee
(where $\Lambda$ is the Prigogine operator which implements the boost) and if
\be
\frac{d \Omega}{dT_E}\leq 0
\ee

where $T_E$ is the Euclidean time in the observers frame, then the observer is accelerating wrt to the background, and NOT in free fall.

\section{Bifurcate Horizons}
To posit the same derivation in a black hole, we begin with an eternal non-rotating black hole. The Kruskal extension of this space-time, has the same causal structure as that of the Rindler observer, two asymptotic regions demarcated by a pair of horizons. Time flows forward in one asymptotics, and backwards in the other. Exactly in a way similar to the Rindler space-time, the ignoring of qft in one asymptotic, as implemented using a Prigogine transformation leads to thermal behavior of the quantum field in the other. However, unlike the Minkowski Diamond example, the time reversed modes of the scalar field in the Kruskal extension donot exist. Thus causal structure of the black hole imposes new boundary conditions which ensure that the Kruskal space-time has entropy.

\subsection{QFT on the Kruskal extension}
The Kruskal extension of the non-rotating black hole metric discovered by K. Schwarzschild has two asymptotic regions, separated by two null surfaces, the past and future horizons \cite{wald}. The Penrose diagram for this is shown in the figure, and the two asymptotics are labeled as I and II. 
This is analogous to the Minkowski causal diamond Penrose diagram, and represents the maximal extension of the Schwarzschild space-time. 
Thus one can directly see why this physical example is similar to the Minkowski-Rindler discussion of the thermalization.

 %horizons are regions of the space-time where the time like Killing vector has vanishing norm. This Killing vector has a similar role to the `boost isometry' of the Rindler space-time. 

We begin with the Schwarzschild metric and derive the Kruskal extension. The Schwarzschild metric is given in $(t,r,\theta,\phi)$ coordinates as
\be
ds^2= - \left(1-\frac{2GM}{r}\right) dt^2 + \left(1-\frac{2GM}{r}\right)^{-1} dr^2 + r^2 d\Omega
\ee

$d\Omega$ is the angular metric, $r$ is the radial coordinate, $M$ the mass of the metric.
The vector $ n^{\mu}\equiv\left( \frac{\partial}{\partial t}\right)^{\mu}$ obviously generates isometries, and is thus a Killing vector. The norm of the Killing vector $ n^{\mu}n_{\mu}= g_{tt}= \left(1- \frac{2GM}{r}\right)$ is 0 at $r=2GM$ and this locates the horizons. 
It is interesting that near the horizon, as $r\approx 2GM$ the metric is Rindler, and thus the discussion smoothly connects with the previous section about the Rindler observer. Given $\rho= r-2GM$, the metric can be re-written as
\bea
ds^2 & = & -\frac{1}{r} \left(r - 2GM\right) dt^2 + \frac{ r} {\left( r - 2GM \right)}  \ dr^2 + r^2 d\Omega\\ 
ds^2 & = & - \frac{\rho}{2GM} dt^2 +  \frac{2GM}{\rho} d \rho^2 + 4G^2 M^2 d\Omega
\eea

And thus the scalar wave function is easy to map to the wave function of (\ref{eqn:sol}). As is evident there as $\rho\approx 0$, the scalar wave function is of the form (\ref{eqn:near}).
Thus the observer at the horizon is `accelerating' with an acceleration at $ a= 1/2GM$. The `Minkowski' coordinates at the horizon is obtained by defining the Kruskal extension of the Schwarzschild space-time. The horizon is smooth in these coordinates.

The scalar wave-function can be solved in these coordinates:
\be
-\frac{2GM}{\rho} \partial_t^2 \phi + \frac{1}{2GM} \partial_{\rho} \left(\rho\partial_{\rho} \phi\right) + \frac{1}{4 G^2 M^2} \frac{1}{\sin\theta}\partial_{\theta}\left(\sin\theta \partial_{\theta}\phi\right) + \frac{1}{4 G^2 M^2 \sin^2\theta} \partial^2_{\phi}\phi=0
\ee
The above equation can be solved using the ansatz
\be
\phi= e^{- i \ \omega\  t}\chi Y_{lm}
\ee
where we have taken $k^2= l(l+1)$, the radial equation reduces to:
\be
\rho^2 \frac{d^2 \chi}{d \rho^2}+ \rho \frac{d\chi}{d \rho} + \left[ 4 (GM)^2 \omega^2 - \frac{k^2}{2 GM} \rho\right] \chi=0
\ee
This is solved by 
\be
\chi= A_I J_{4 i GM \omega} \left( 2 i k' \sqrt{\rho}\right) + B_I Y_{4 i GM \omega} \left( 2 i k' \sqrt{\rho}\right)
\ee
  with $k'= \sqrt{2/GM} k$.
Again near the horizon as in (\ref{eqn:near}) the solution is approximated as
\be
\phi ^I \approx A'_I \  e^{-i \ \omega \ t + i  \ 2 G M \omega \ \ln\rho}\  Y_{lm} + B'_I \ e^{-i \ \omega \ t - i \ 2 GM \omega \ \ln \rho}\  Y_{lm} + c.c. 
\label{eqn:wave}
\ee

where we have absorbed some of the pre-factors from the Bessel function in the $A' $ and the $B'$ constants.
Note we introduced the notation $\phi^I$ here to indicate that there are two such Rindler universes one on each side of the horizon.
The other side of the horizon will have
\be
\phi^{II} \approx  A'_{II} \  e^{- i \ \omega\  \tilde t + i \  2GM \omega \ln\rho}\  Y_{lm} + B'_{II} \ e^{- i \ \omega \ \tilde t - i \ 2GM \omega \ln \rho} \ Y_{lm} + c.c.
\ee

However, this time flows backwards, and thus we use the change $\tilde t=- t$ before transforming this to the U, V coordinates:
\be
\phi^{II} \approx A'_{II}\  e^{ i \ \omega  \ t + i \ 2GM \omega \ln \rho}\  Y_{lm}  + B'_{II} \ e^{i \ \omega \ t -i \ 2GM \omega \ln \rho}\  Y_{lm} + c.c.
\ee

%coordinate lines of the light like geodesics, in a conformal Penrose diagram of this, the rays are at 45 degree angles.

To define the wave function in coordinates across the horizon, we use the following

\small{
\bea
u &=&  t -r - 2GM \ln \left(\frac{r}{2GM}-1\right) \approx t - 2GM \ln (r - 2GM) -r + 2GM \ln (2GM)\approx t- 2GM \ln \rho -r + 2GM \ln (2GM) \nn \\
v&=& t + r + 2GM \ln \left(\frac{r}{2GM} - 1 \right) \approx t + 2GM \ln (r -2GM) + r + 2GM \ln (2GM) \approx t + 2GM \ln\rho + r + 2GM \ln (2GM) \nn
\eea}

The Kruskal coordinates are defined by
\bea
U &= & - e^{-u/4GM}\\
V &=& e^{v/4GM}
\eea

%The Kruskal extension is obtained, in terms of new coordinates $U, V$ defined thus

%\bea
%UV & = &  e^{r/2GM} \left(\frac{2GM}{r}-1\right) \nn \\
%\frac{V}{U} & =  & e^{t/2GM}
%\eea

\begin{figure}
\includegraphics[scale=0.7]{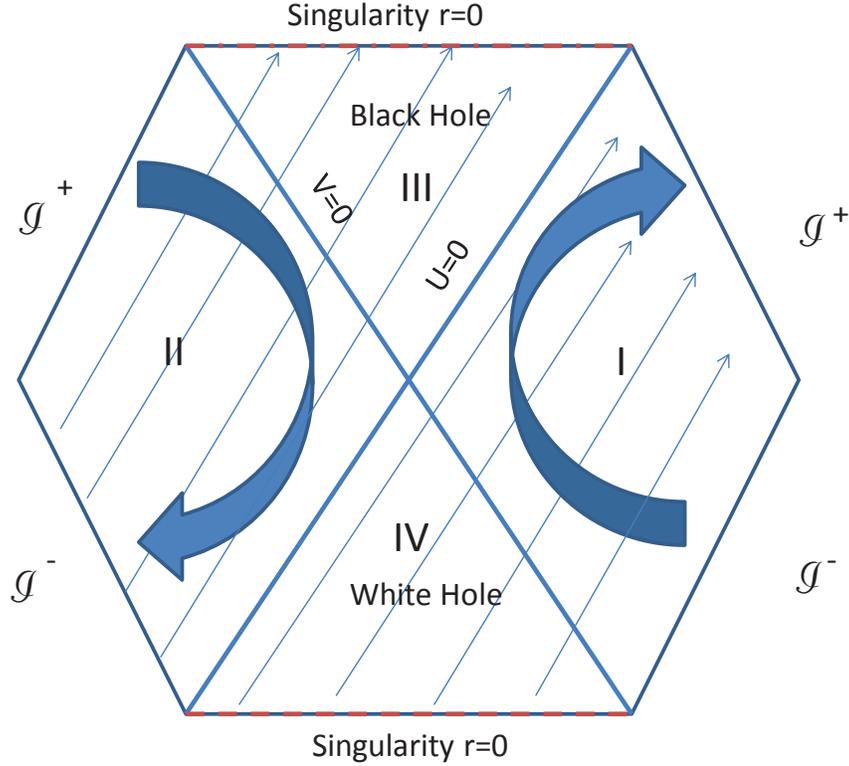}
\caption{Kruskal Space-time}
\end{figure}
 
The above coordinates simulate the lightcone `Minkowski coordinates', and represent light-like geodesics.  The particles are in free fall along these coordinates. We have to find the normal modes of the scalar field, and achieve the transformation to the Rindler versions.
The metric in Kruskal coordinates is.

\be
ds^2 = -32 (GM)^3\frac{ e^{-r/2GM}}{r} dU dV + r^2 d\Omega
\label{eqn:kruskal}
\ee

In this coordinate the metric is smooth at $r=2GM$, the metric in the $U,V$ sector is conformal to Minkowski space,  and the location of the past horizon is V=0 and future horizon is U=0.  

Exactly in analogy to the Rindler observers's space-time, the U=0 and the V=0 axes separate the entire space-time into four regions labeled as I, II, III, IV. The regions I and II have the asymptotics, $r\rightarrow \infty$ and are causally disconnected from each other. The Regions
III and IV include the $r=0$ curvature singularities. These are regions of the `black hole' and the `white hole'. Matter falls into the black hole region III disappearing into the singularity and matter is emergent from the singularity escaping to the outside from the white hole region IV.
The `white hole' and the `black hole' regions require that the scalar field flows in one direction only in these regions, and thus the time reversed modes used in the example of Minkowski space $\ovl a$ cannot exist in these regions .

We begin with the QFT as in the previous case defined on one of the horizons, taken as an almost Cauchy Surface: in this example the 
V=0 surface, or the past horizon.

Exactly as in the Rindler example, we take the almost Cauchy surface $V=0$ and write a typical quantum field, expanded in the modes of the given regions I and II and transform to the Kruskal coordinates which `connects' the two regions. 
Thus

\bea
\Phi & = & \int d\om \  \  [ \phi^{I} \  a^{I} + \phi^{ I*} \ a ^{ \dag \  I}]   \  \   \  \    \   \   \   \   \   \   \   U<0 \\
&=& \int d \om \  \  [\phi ^{II} \ a^{II} + \phi^{II*} \ a ^{\dag \ II} ] \ \  \  \   \   \   \  \  \  \  \  \   \  \  U>0
\eea

We then obtain the above as written in terms of the wave functions of the modes of the (\ref{eqn:kruskal}) metric. To determine the Kruskal modes, we use the approximation that as the waves near the horizon, the geometric optics approximation can be used to estimate the wave function. Let us describe what this is for a generic space-time: Given the scalar wave equation we assume that the wave function is of the form $e^{i \ S}$, as in this approximation, the phase of the wave function dominates:

\bea
\frac{1}{\sqrt{-g}}\partial_{\m}\left(\sqrt{-g}  \  g^{\m \n}\partial_{\n} \ \ e^{ \ i S}\right) & = & 0\\
\frac{i}{\sqrt{-g}} \ \partial_{\m}\left(\sqrt{-g} \ g^{\m \n} \partial _{\n}  \ S\right) - g^{\m \n} \ \partial_{\m} S \partial_{\n} S &=& 0 
\eea
 
The imaginary first term is neglected in the WKB approximation (being linear in S) and we are left with the second term, which has the form of a null norm $n^{\mu}n_{\mu}=0$ where $\partial_{\mu} S=n_{\mu}$ is a null vector, the $S$ can be identified as proportional to the affine parameter U along the null surface $V=0$. Thus, in the WKB or geometric optics approximation the Kruskal modes at the horizon are
$ e^{i \  \omega  \ U} Y_{lm} (\theta,\phi)$.

As is evident the wave function in the Rindler metric is of the for given as  (\ref{eqn:wave}).
Thus
\bea
\phi^I &= & \ \int \  d\om \ [ A'_I \  e^{i \om \ u}{e^{+r + 2GM \ln 2GM}} + B'_I  \  e^{i \om \ v}{e^{-r - 2GM \ln 2GM}} + c.c.]\\
&=& \  \int d \om \  [ A'_I f(r) \left(- U\right)^{4GM i \omega} +B'_I g(r) (V)^{ 4 GM i \om} + c.c. ...]
\eea
for $U<0$, where $f(r)= e^{r + 2GM \ln (2GM)}$, as in the metric, and 
$g(r)= e^{-r -2GM \ln 2GM}$.

Similarly $\phi^{II}$ is obtained as
\be
\phi^{II}= \int \ d\om \left[ A'_{II}  (V)^{4GM i \om}  g(r) + B'_{II} (-U)^{4GM i \om} f(r) + c.c. \right]
\ee

In the near horizon limit, we simply put $r=2GM$ and absorb the $f(2GM)$ into the constants which are then re-defined as $\tilde{A}_I, \tilde{B}_I$ etc. 
Thus, using the Fourier transforms of the modes $\phi^I$ and $\phi^{II}$ and the c.c. wrt the modes $e^{i \ \om \ U}$ and $ e^{-i \ \om\  U}$, one finds that
\bea
\Phi &= &  \int\  d \om\  d \om' \left[ \tilde{A}_I  \frac{1}{(i \omega')^{4 GM i \om +1} }\Gamma ( 4 GM i \omega +1) Y_{lm}(\theta, \phi)  \ a_{\om \ l \ m \ I } \ e^{i \ \om' \ U}   \right. \nn \\ &&  + \tilde{A}_I  \frac{1}{(-i \om')^{4GM i \om +1}} \Gamma(4 GM i \om +1) Y_{lm} (\theta, \phi) \ a_{ \om\  l\ m \ I} \  e^{- i \ \om' \ U} + c.c. \nn \\ 
&& + e^{-4 GM \pi \om} \tilde{B}_{II} \frac{1}{(- i \omega')^{-4 GM i \om +1}}  \Gamma( 1 - 4GM i \om) Y_{lm} (\theta,\phi) \ a_{ \om\  l \ m \ II } e^{- i \ \om' \ U}  \nn \\ &&  + e^{-4 GM \pi \om} \tilde{B}_{II}  \frac{1}{(i \omega') ^{1- 4GM i \om}} \Gamma(1 - 4 GM i \om) Y_{lm} (\theta, \phi) \  a_{ \om \ l \ m  \ II}  e^{ \ i \ \om' U} + c.c. \left. \right]
\eea

We then collect the terms proportional to  $e^{- i \om' U}$ and identify the operators as two different types of annihilation operators:
 \bea
{\ovr{a}_ {\om lm}} & = & a_{ \om l m \ I}  + e^{-4\pi GM} a_{ \om l -m \ I}^{\dag}\\
\ovl{a}_{\om lm} &=& a_{\om l m \ II} + e^{- 4\pi GM} a_{\om l -m I}^{\dag}
\eea
and their complex conjugates.

\be
|0_{RI}>= \prod_{\om}e^{-\tilde{z} \ovr{a}^{\dag}_{\om l m} \ovl{a}^{\dag}_{\om l m}}|0>
\ee

where $\tilde{z}= e^{-4 \pi GM}$.
Exactly as in the example of Minkowski $\rightarrow$ Rindler transformations, the dictionary can be built, of the map of the operators and density matrices. In this case, however we make the interesting observation that due to the presence of the white whole and the black hole regions, on the V=0 surface, the $\ovl{a}$ modes do not exist. Thus, when we define the map from the Rindler vaccum to the Minkowski vacuum, or the map from the Schwarzschild observer to the Kruskal observer, only one set of time forward modes are picked up. This induces a thermalization which does not happen in Minkowski space-time.

%Thus, using a Lyapunov function,
%\be
%\Omega= {\rm Tr}(\rho^{\dag}\rho)
%\ee
%such that $\rho= |0_{K}><0_{K}|$, and the system is evolved in Minkowski time, with a Hamiltonian $H= \int d\om \ovr{a}_{\om}^{\dag} %\ovr{a}_{\om}$ such that only the forward modes are created, we obtain a Lyapunov function as in the example of the Rindler modes.
%Note that if we have the entire Rindler Hamiltonian
%$H= \int \omega a_{I}^{\dag} a_{I}+ a_{II}^{\dag} a_{II}$, and transform to the Kruskal modes,
%$H= \int \omega \ovr{a}^{\dag}\ovr{a} - \ovl{a}^{\dag}\ovl{a}$, and exlusion of one set truncates the system.e
%The Kruskal vacuum corresponds to the Hartle-Hawking vacuum, and we have a very transparent reason why the system is thermalized.
 
As the $\ovl{a}$ modes donot exist, we can only deal with density matrices and trace over the $\ovl{a}$ modes of the system.
In the usual framework of the `tracing mechanism' for determining entropy of quantum states;
As
\be
|0_{RI}>=\prod_{\om} e^{-\tilde{z} \ovr{a}^{\dag}_{\om l m} \ovl{a}^{\dag}_{\om l m}}|\ovr{0}>\otimes|\ovl{0}>=\prod_{\om}\sum_{n} (-\tilde{z})^n |\ovr{n}>\otimes|\ovl{n}>
\label{eqn:minvacc}
\ee

The system is in the direct product space of the forward and time reversed modes. Tracing over one set, as in the example of the black hole, due to causal structure of background, one obtains a density matrix which has entropy.
\be
\rho_R= {\rm Tr} ~ |0_{RI}><0_{RI}|= \prod_{\om}\sum_{n} (-\tilde{z})^{2n}|\ovr{n}><\ovr{n}|
\ee

This is a thermalized density matrix \cite{unrhwald}. Note that in this example, the boundary conditions ensure that $\ovl{a}_{\om l m}$ modes do not exist. This makes this state thermalized ab initio. One doesn't have to use Rindler time to find the entropic behavior.

The other derivations of the section on Minkowski - Rindler space-time would be exactly the same for the Kruskal-Rindler transformations too: e.g.
\be
a_{\om l m  \  I} = U \ovr{a}_{\om \ l\  m}U^{-1}  \    \      \     \     \    \   a_{\om \ l  m \ II}= U\ovl{a}_{\om l m} U^{-1}
\ee

This permits the evaluation of $d \rho_R/dt$ as in  (\ref{eqn:timev}), and the anti-Hermitian term is similar in origin for the Hawking radiation as observed in \cite{adg12,adg2}, when we trace over the $a_{II}$ modes.
In a collapsing situation, both the $\ovl{a}$ and the $a_{II}$ modes donot exist.

\section{Postulating QFT for accelerated frames}
Based on the observations of the above sections, I try to postulate some generic rules for QFT to include accelerated frames/curved space-times.

\noindent
1. QFT is defined generically as the direct product space of two Hilbert spaces which comprises of modes flowing forward and backward in time $H= H_{\rm I} \otimes H_{\rm II}$.\\
2. In case of accelerated observers, the two Hilbert spaces (time forward and time backwards) are defined in space-time regions separated by a bifurcate horizon.\\
3. The star-Unitary operator $\Lambda(L)$ and the star-Hermitian operators $\Phi$ are important for QFT in accelerated frames. The $\Lambda(L)$ operators map the density matrices of the inertial frame to the accelerated observers frame. The $\Lambda$ operator maps to Minkowski density matrices to density matrices in $H_I$ and a second $\Lambda$ operator maps the Minkowski desnity matrices to density matrices of $H_{II}$. Irreversible physics emerges if we keep one $\Lambda$ operator, and one $H_I$. The time evolution in the accelerated frame is governed by $\Phi$ a star-Hermitian operator which allows pure states to evolve into mixed states.\\
4. The time evolution in the truncated Hilbert space ($H_I$ or $H_{II}$) might be non-Unitary, with plausible entropy creation.\\
%6. Thermalisation can also arise due to boundary conditions necessitated by causal structure of the space-time. In this case too, the presence of the Prigogine operators $\Lambda,\Phi$ make the QFT description complete.\\ 

%\be
%\int_{-\infty}^{0} A_I  a_{I} f (r) (-U)^{4 GM i \om} e^{ i \om U} \ dU + c.c.
%\ee

%[To be Filled]

 %\subsection{Generic Bifurcate Horizons}

%The above description of obtaining a thermalized state by Prigogine transforming a vacuum state can be extended to the example of 

%\section{The Algebraic Formulation and the Hadamard Vacuum}

\section{Conclusions}

Thus we showed that the Bogoliubov transformation which maps Minkowski vacuum to the vacuum of a Rindler observer is a Prigogine transformation. The existence of a Lyapunov function shows that the Rindler observer perceives irreversibility.  The Rindler observer sees one half of a direct product space, one in which time flows forward and the other in which time flows backward.  The Prigogine transformation chooses one of the spaces with one time flow, and thus physics for a Rindler observer is irreversible. The transformation which takes the Minkowski density matrix to the other Rindler Hilbert space's density matrix, in this example happens to be identical. However, it can be labeled as star-Unitary operator 2. Thus there are two Prigogine transformations, one which maps to Hilbert space in which time is flowing forward, and the other to the Hilbert space in which time flows backward. In the Rindler example there is no loss of information, entire information is contained in the direct product of two Hilbert spaces ($H_I \otimes H_{II}$). Mapping to one $ H_I$ or $H_{II}$ gives rise to irreversible physics.  

The tracing mechanism, which traces over one of the $H_I$ or $H_{II}$ basis states when the Minkowski state is written in the direct product Hilbert space basis is a different from the Prigogine map discussed here. The star-Unitary operators `break' the time reversal symmetry of the direct product space $H_I\otimes H_{II}$, by projecting the Minkowski density matrices defined in the direct product space, to one of the Rindler Hilbert spaces. Thus this transformation, is `microscopic' as implemented on the density matrices, and does not involve a tracing mechanism. The end result is though the same, the remaining density matrix/quantum state is written in one of the Hilbert spaces, $H_I$ or $H_{II}$. This is a concrete example of the implementation of the Prigogine formulation, and clarifies the role of the star-Unitary operator in a evidently time reversal symmetric system. There is a breaking of symmetry due to the use of the star-Unitary operators. However, there are two star-Unitary operators for each time direction (in this example the `time directions' are represented by $H_{I}$ and $H_{II}$ Hilbert spaces). This is in agreement with a discussion on the Prigogine formalism \cite{phil}. 

The same formalism can be used for Bifurcate Killing horizons where the near horizon observer is accelerating with respect to the background metric. However for the black hole, additional boundary conditions due to presence of white whole and black hole regions cause thermalisation not present in the Minkowski-Rindler example.

Our eventual aim is to develop a generic QFT for accelerated observers, including that which is valid in arbitrary curved space-times. To define quantum field theory in space-times without Killing vectors, one needs to use the formalism of Algebraic Quantum Field Theory and Hadamard condition \cite{kay}, and this is work in progress \cite{adg1}. One also has to address the problems of interaction of the QFT and renormalization in accelerating frames as in \cite{accel}.


\begin{thebibliography}{99}
\bibitem{qft} P. A. M. Dirac, {\it The Quantum Theory of the Emission and Absorption of Radiation,} Proceedings of the Royal Society of London, Series A, Vol.114, (1927) 243.  
\bibitem{qft1} J. D. Bjorken and S. Drell, Relativistic Quantum Mechanics, McGraw-Hill College (1965).
\bibitem{qgr} S. Carlip, D. Chiou, W. Ni, R. Woodard, {\it Quantum Gravity: A Brief History of Ideas and Some Prospects} arXiv:1507.08194 [gr-qc]. and references therein. J. Ambjorn, A. Goerlich, J. Jurkiewicz, R. Loll, {\it Nonperturbative Quantum Gravity} Physics Reports 519 (2012) 127.
 \bibitem{prig}  I. Prigogine, {\it From Being to Becoming: Time and Complexity in the Physical Sciences} WH Freeman \& Co (1980).
\bibitem{wald} R. M. Wald, {\it General Relativity} University of Chicago Press (1984). 
\bibitem{marolf} D. Marolf, {\it Notes on Relativity and Cosmology} (2003).
\bibitem{unruh} W. G. Unruh, Phys. Rev. {\bf D 14} (1976) 870. 
\bibitem{unrhwald} W. G. Unruh, R. J. Wald Phys. Rev. {\bf D32} (1985) 831.
\bibitem{bd1} N. D. Birrell and P. C. W. Davies, {\it Quantum Fields in Curved Space}, Cambridge Monographs in Mathematical Physics, (1984).
\bibitem{rindl} T. Padmanabhan, Mod. Phys. Let. {\bf  A17} (2002) 1147.
\bibitem{rindl2} R. Laflamme, Phys. Lett. {\bf B 196} 449.
\bibitem{accel} N. Sanchez, Phys. Rev. {\bf D24} (1981) 2100. N. Sanchez and B. F.  Whiting Phys. Rev. {\bf D 34} 1056. 
\bibitem{accel2} T. Padmanabhan Phys. Rev. Lett. {\bf 64} (1990) 2471.
\bibitem{boltz} L. Boltzmann, Sitzungsberichte Akademie der Wissenschaften 66 (1872) 275.
\bibitem{michel} M. DeHaan and C. D. George, Prog. Theor. Phys. Vol. 109  (2003) 881.
\bibitem{lyapunov} A. M. Lyapunov, {\it The General Problem of the stability of motion}, Taylor and Francis, London (1992).  
\bibitem{thooft} G. 't Hooft Int.J. Mod. Phys. {\bf A11} (1996)  4623.
\bibitem{adg12} A. Dasgupta, J. Mod. Phys. {\bf 3} (2012) 1289. 
\bibitem{adg2} A. Dasgupta SIGMA {\bf 9} (2013) 013.
\bibitem{kay} B.S. Kay and R.M. Wald Phys. Rept. {\bf 207} (1991) 49.
\bibitem{adg1} A. Dasgupta, {\it Accelerated observers and Algebraic Quantum Field Theory} to appear.
\bibitem{phil} V. Karakostas, Phil. Sc. {\bf 63} (1996) 374.
\end{thebibliography}
\end{document}